\documentclass[namedreferences]{solarphysics}

\usepackage[hyperref,optionalrh,natbib]{spr-sola-addons} 
\usepackage{graphicx}        
\usepackage{color}           
\usepackage{breakurl}        

\begin{document}

\begin{article}

\begin{opening}

\title{Objective Image Quality Assessment for High Resolution Photospheric Images by Median Filter Gradient Similarity  }

\author{Hui~\surname{Deng}$^{1}$\sep
	    Dandan~\surname{Zhang}$^{1}$\sep
        Tianyu~\surname{Wang}$^{1}$\sep
        Kaifan~\surname{Ji}$^{1}$\sep
        Feng~\surname{Wang*}$^{12}$\sep
        Zhong~\surname{Liu}$^{2}$\sep
        Yongyuan~\surname{Xiang}$^{2}$\sep
        Zhenyu~\surname{Jin}$^{2}$\sep
        Wenda~\surname{Cao}$^{3}$
       }
\runningauthor{Deng et al.}
\runningtitle{Median Filter Gradient Similarity}

   \institute{$^{1}$ Computer Technology Key Lab of Yunnan Province, Kunming University of Science and Technology\\
                     email: \url{dh@cnlab.net} email: \url{wangfeng@acm.org, corresponding author}\\
              $^{2}$ Yunnan Observatories, Chinese Academy of Sciences \\
              $^{3}$ New Jersey Institute of Technology, Newark, NJ, United States \\
             }

\begin{abstract}
All next generation ground-based and space-based solar telescopes require a good quality assessment metric in order to evaluate their imaging performance. In this paper, a new image quality metric, the median filter gradient similarity (MFGS) is proposed for photospheric images. MFGS is a no-reference/blind objective image quality metric (IQM) by a measurement result between 0 and 1 and has been performed on short-exposure photospheric images captured by the {\it New Vacuum Solar Telescope} (NVST) of the Fuxian Solar Observatory and by the {\it Solar Optical Telescope} (SOT) onboard the {\it Hinode} satellite, respectively. The results show that: (1)the measured value of MFGS changes monotonically from 1 to 0 with degradation of image quality; (2)there exists a linear correlation between the measured values of MFGS and root-mean-square-contrast (RMS-contrast) of granulation; (3)MFGS is less affected by the image contents than the granular RMS-contrast. Overall, MFGS is a good alternative for the quality assessment of photospheric images.
\end{abstract}

\end{opening}


\section{Introduction}
\label{S-Introduction}
The quality of the images captured by ground-based telescopes is heavily limited by the Earth's atmospheric turbulence which is commonly termed as ``seeing''. For telescopes without Adaptive Optics (AO) systems, to alleviate the image degradation induced by the atmospheric turbulence and to achieve higher angular resolution, post-facto reconstruction techniques ({\it e.g.} speckle image reconstruction) are widely used. Even for those telescopes with AO systems, the observed images are often processed {\it post-facto} to further reduce residual aberrations and to achieve the diffraction limit of the telescope over a larger field of view (FOV). Under such circumstances, a proper objective image quality metric (IQM) is required to evaluate the quality of the short-exposure images (frames) captured by high-speed cameras. The frames with  higher values of such a metric might be selected for the subsequent reconstruction process. Namely, an improved IQM will be utilized to streamline the {\it post-facto} image processing.

Based on the amount of information required, objective IQMs can be classified into full-reference (FR), no-reference (NR)/blind and reduced-reference (RR) ones. FR IQMs require the entire reference image to be available, usually without blemishes. The mean squared error (MSE) is the simplest FR IQM, computed by averaging the squared intensity differences of distorted and reference image pixels. The structural similarity (SSIM) proposed by \citet{Wang2004} is one of the most widely used FR IQMs because its measurement result is close to the result given by human visual system. In many practical applications ({\it e.g.} astronomical observations), the reference image is not well-defined, so that NR/blind IQMs are preferred. Some NR IQMs have been studied or used in astronomical observation and image processing applications. The sharpness of an image measured after high pass filtering was used for real-time frame selection by the {\it Swedish Vacuum Solar Telescope}~\citep{Scharmer1989}. \citet{Bos2012} compared the effect of sharpness and entropy in tuning the inverse filter used for amplitude recovery in a speckle imaging system. The Fisher information is a measure of disorder and was used by \citet{Zhangsijiong2006} for searching lucky images. The Fourier amplitude was used by \citet{Garrel2012} for frame selection.

For the quality assessment of high resolution photospheric images, the most commonly used is the root-mean-square-contrast (RMS-contrast)~\citep{Denker2005,Denker2007,Danilovic2008,Scharmer2010}. When the measure of RMS-contrast is only applied to the quiet-Sun region of a photospheric image, it is called the granular RMS-contrast. One shortcoming of the granular RMS-contrast is that it depends on the wavelength used \citep{Albreg1977,Ricort1979} and also on the distance from the disk center \citep{Cuberes2000}; with the increasing heliocentric angle or observing wavelength, the granular RMS-contrast decreases (see Figures 1 and 2 in~\citet{Albreg1977}). Moreover, the RMS-contrast value of an image depends strongly on the structural contents of the assessed image. When dark features (pores or sunspots) are moving in and out of the field of view (FOV), the value of RMS-contrast must be interpreted carefully.

In this paper, we present the median filter gradient similarity (MFGS), a new NR objective IQM. As you see below, MFGS shows advantages in measuring the quality of photospheric images over {\it e.g.} RMS-contrast. The rest of this paper is organized as follows. In Section \ref{S-mfgs}, we introduce MFGS. In Section \ref{S-Experiment}, the performance of MFGS on photospheric images is validated. In Section \ref{S-Comparison}, MFGS and RMS-contrast are compared by looking at their mutual correlation and their dependences on the image contents. In Section \ref{S-ConclusionDiscussions}, discussion and conclusion are given.

\section{Median Filter Gradient Similarity} 
\label{S-mfgs}
\subsection{Principle of MFGS}

\label{sub-mfgsPrinciple}
The MFGS metric originated from both the SSIM assessment method \citep{Wang2004} and the NR perceptual blur metric \citep{Crete2007}. SSIM was designed to improve the traditional algorithms like peak signal-to-noise ratio and mean squared error, which have been proven to be inconsistent with human eye perception. The main objective of the SSIM metric is to evaluate the similarity between the assessed image and its reference one in terms of luminance, contrast, and structure. However, SSIM is a FR metric and cannot be directly used in astronomical applications. The NR perceptual blur metric is based on the property that it is difficult for a human being to perceive differences between a blurred image and its re-blurred one.

Since the observation of small-scale structures in the photosphere is one of the main scientific goals for solar telescopes, structural information is a key factor that we must consider when proposing a new IQM. In this aspect, the pre-processing of a raw image by a noise removal filter enhances the effectiveness of IQM when the image quality is low.

\begin{figure}    
\centerline{\includegraphics[width=1\textwidth]{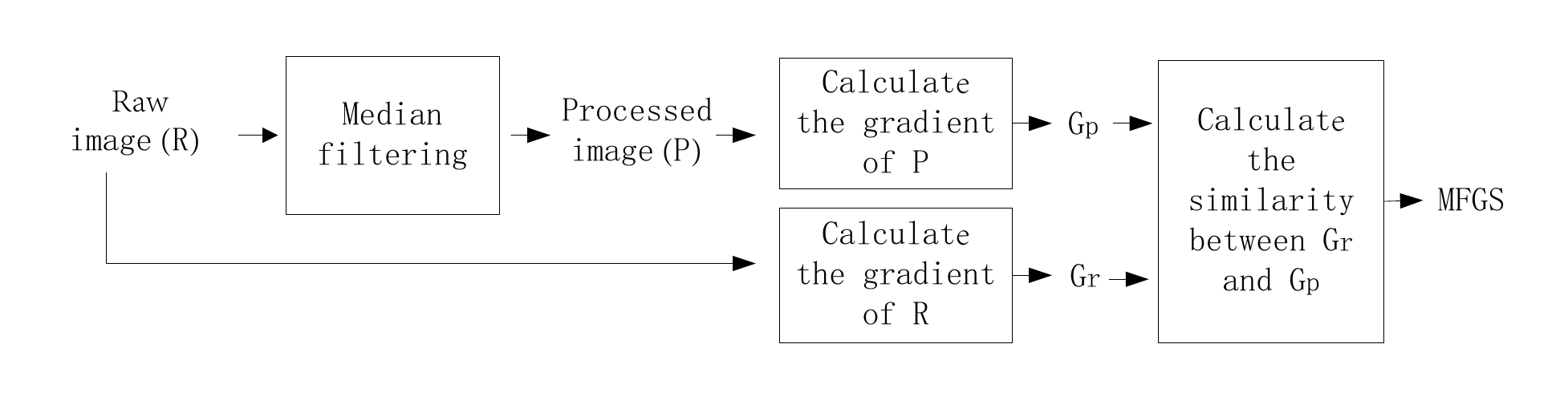}}
\caption{%
The flowchart of the MFGS metric.
}
\label{F-MFGS_2a}
\end{figure}

Based on these considerations, we propose MFGS, a new image quality assessment metric for photospheric images. The basic process of MFGS consists of (1) filtering the raw image (image under evaluation) to obtain a processed image (reference image); (2) calculating the gradients of the raw and the processed images, respectively; (3) calculating the similarity between the two gradients. The flowchart is shown in Figure~\ref{F-MFGS_2a}.

\subsection{Algorithm of MFGS} \label{sub-mfgschart}
As was stated above, MFGS contains three steps, {\it i.e.}, filtering the raw image, calculation of the gradient, and calculation of the similarity. They are described below in detail.

\subsubsection{Step 1 - Filtering the Raw Image with a Median Filter}
A median filter, a nonlinear operation, is more effective than convolution in order to reduce noise and preserve edges simultaneously~\citep{Dbrown1984,Lin1988,Tsun1994,Wang1999,AriasCastro2009}. A 3~$\times$~3 median filter is applied to the raw image ($R$) to obtain the processed image ($P$):
\begin{equation}
P=M\left(R, [3~3]\right),
\label{eq-filtering}
\end{equation}
where $M$ is the two-dimensional (2-D) median filtering operation, and [3 3] is a 3~$\times$~3 local window.

\subsubsection{Step 2 - Calculating the Gradient Value of $R$ and $P$}
Structural information in the images is represented by the gradient, which is an intuitive and simple measure of structural distortion and is widely used in objective image quality assessment \citep{Asatryan2009,Liang2010,Kim2010,Zhu2012}. Two gradient values, $G_{\rm r}$ for the raw image and $G_{\rm p}$ for the processed image, are obtained as

\begin{equation}
  G_{\rm r}=\sum|D(R)|,
  \label{eq-gradient1}
\end{equation}

\begin{equation}
  G_p=\sum|D(P)|,
  \label{eq-gradient2}
\end{equation}
where $D$ is the gradient operator of difference ($\left[ \begin{array}{cc}
-1 & 1 \end{array} \right]$).

\subsubsection{Step 3 - Calculating the Similarity between $G_{\rm r}$ and $G_{\rm p}$}
The following equation, which has been widely used in similarity-based image quality assessments~\citep{Wang2002,Wang2004,Chen2006,Rehman2012,Zhu2012}, is adopted to calculate the similarity between \emph{R} and \emph{P},
\begin{equation}
\textrm{MFGS}=\left(2G_{\rm p}G_{\rm r}\right)/\left(G_{\rm p}^2+G_{\rm r}^2\right).
\label{eq-mfgs}
\end{equation}
Obviously, MFGS takes a value from 0 to 1. The quality of the image is better if its MFGS value is higher; a perfect image is represented by MFGS = 1.

\section{Performance Evaluation}
\label{S-Experiment}
\subsection{Sample Data}
\label{S-Experimentdata}
Sample images were taken with the {\it New Vacuum Solar Telescope} (NVST; \citet{Liu2011}; \citet{Liu2014}). NVST is a vacuum solar telescope with 1 m aperture that aims to observe fine structures on the Sun. It is the main observation facility of the Fuxian Solar Observatory (FSO) located at $24^{\circ}34'48''$N, $102^{\circ}57'01''$E, on the northeast side of the Fuxian Lake, Yunnan, China. Its high resolution imaging system consists of one chromosphere channel (H$\alpha$, 6563 {\AA}) and two photosphere channels (TiO band, 7058{\AA}, and G-band, 4300{\AA}).

The sample photospheric images were obtained at the wavelength of 7058 {\AA} by a high-speed CMOS camera (10-15 frames per second, about 1 millisecond exposure time). The FOV was $102'' \times 86''$ with an image scale of 0.04$''$/pixel (2560 $\times$ 2160 pixels). All the images were taken without AO, and basic image corrections (dark subtraction and flat fielding) were made on them.

Four representative data sets taken in September and October 2012 were selected from a huge amount of observational data sets. The four data sets include two active regions, NOAA 11575 and 11598, which represented observations with different quality levels. Each set consists of 200 short-exposure frames taken with a cadence of about 15 s. The seeing parameter $r_{0}$ of each set was calculated by the method developed by~\citet{von1984}. The four sets are labeled as ``A'', ``B'', ``C'', and ``D'' in the descending order of $r_{0}$ values (Table~\ref{T-dataset}).

In addition, a G-band image taken with the {\it Solar Optical Telescope} (SOT; \citet{SOT}) onboard the {\it Hinode} satellite on 28 February 2007 was used to further confirm the performance of MFGS on images without atmospheric turbulence. The image of perfect quality (see ``e'' in Figure~\ref{mfgs-High}) was taken with an exposure time of 51 milliseconds. The FOV was $56'' \times 56''$ with an image scale of $0.05''$/pixel.

\begin{table}
\caption{%
Four sets of sample images taken with NVST. The seeing parameter $r_{0}$ was calculated by the method developed by~\citet{von1984}.
}
\label{T-dataset}
\begin{tabular}{cccccccc}
\hline
Dataset & Start time (UT) & Active region number & Location & $r_{0}$ (cm) \\
\hline
A & 04:51:11, 26 Sep. 2012 & NOAA 11575 & W17N08   & 12.1\\
B & 02:15:02, 29 Oct. 2012 & NOAA 11598 & S11W27   & 9.4 \\
C & 04:33:59, 26 Oct. 2012 & NOAA 11598 & S12E17   & 8.5 \\
D & 02:51:31, 29 Oct. 2012 & NOAA 11598 & S11W37   & 8.2 \\
\hline
\end{tabular}
\end{table}

\subsection{Choice of Gradient Operators}
\label{S-gradient operators}
Gradient calculation is the core in the second step of the MFGS algorithm. There are several operators to obtain the gradient approximation, {\it e.g.}, the Sobel $\left[ \begin{array}{ccc}
-1 & 0 & 1 \\
-2 & 0 & 2 \\
-1 & 0 & 1 \end{array} \right]$, Prewitt $\left[ \begin{array}{ccc}
-1 & 0 & 1 \\
-1 & 0 & 1 \\
-1 & 0 & 1 \end{array} \right]$, Roberts $\left[ \begin{array}{cc}
0 & 1 \\
-1 & 0  \end{array} \right]$ and the difference $\left[ \begin{array}{cc}
-1 & 1 \end{array} \right]$ operators. We conducted an experiment to find out which operator is most suitable for MFGS. The four sets listed in Table~\ref{T-dataset} and the image taken with {\it Hinode}/SOT were processed by MFGS, by applying the Sobel, Prewitt, Roberts, and difference operators. The calculated average values of MFGS are listed in Table~\ref{T-Operators}. One can see that the discrimination power of the Roberts and difference operators is superior to the Prewitt and Sobel operators. The SOT image was taken without atmospheric turbulence and its quality should be much better than any short-exposure images taken by the ground-based NVST. However, when performing MFGS with the Prewitt or Sobel operator, the obtained values for the SOT image are very close to those from NVST. Moreover, the difference operator is faster in computation time than the Roberts operator. Therefore, the difference operator is the most suitable one among the examined operators for MFGS.

\begin{table}
\caption{%
The average values of MFGS derived for different datasets by applying different gradient operators.
}
\label{T-Operators}
\begin{tabular}{cccccc}
\hline
            &SOT     &A       &B       &C      &D        \\

\hline
Roberts     &0.991 &0.837 &0.783 &0.751 &0.726     \\
Prewitt     &0.998 &0.984 &0.966 &0.952 &0.938     \\
Sobel       &0.998 &0.981 &0.960 &0.945 &0.930     \\
difference  &0.984 &0.710 &0.616 &0.590 &0.552     \\
\hline
\end{tabular}
\end{table}

\subsection{Performance of MFGS on Sample Data}
\label{Sub-Experimentresults}
\begin{figure}    
\centerline{\includegraphics[width=1\textwidth,clip=]{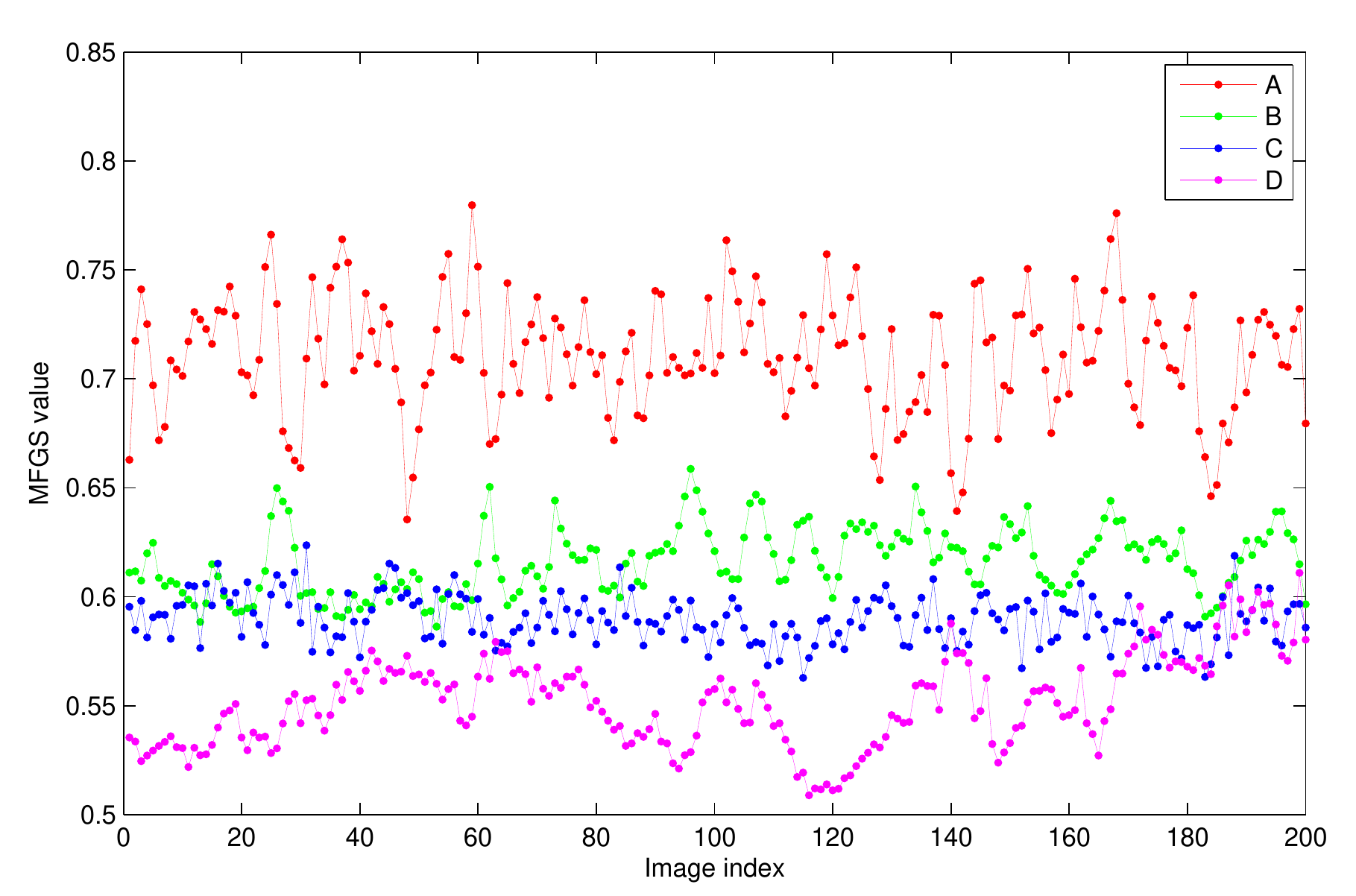}}
\caption{%
Variations in MFGS values versus image index (or time). Each sample point represents one image frame.
}
\label{mfgs-full}
\end{figure}

MFGS was applied to each image frame in the data sets listed in Table~\ref{T-dataset}. Figure~\ref{mfgs-full} shows the variations in the MFGS values versus image index ({\it i.e.}, time). Overall, all of the MFGS values are between 0.5 and 0.8; the average values for sets A, B, C, and D are 0.710, 0.616, 0.590, and 0.552, respectively. These values are consistent in the order with their r$_{0}$ values. In each data set, the MFGS value fluctuates among the frames, which reflects the fluctuation of image quality due to atmospheric turbulence.

\begin{figure}    
\centerline{\includegraphics[width=1\textwidth,clip=]{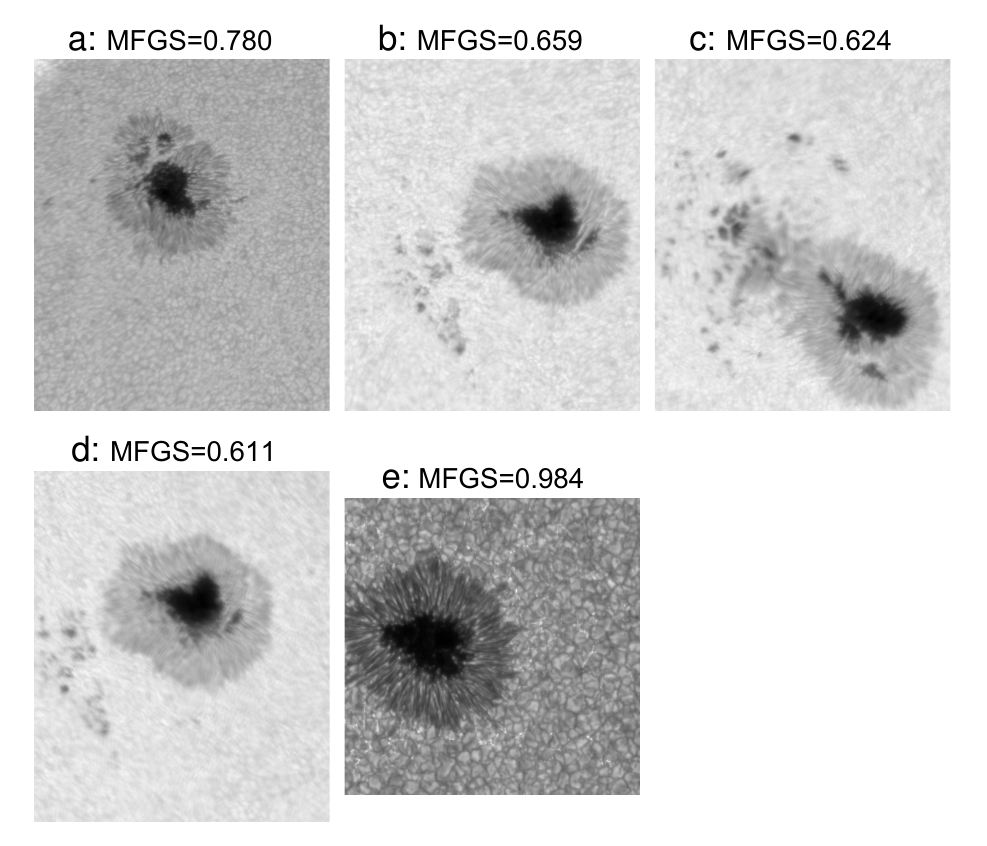}}
\caption{%
The images with the highest MFGS value from data sets A, B, C, and D. The observing wavelength is 7058 {\AA} and the FOV is $102'' \times 86''$. Panel e is the G-band image observed with {\it Hinode}/SOT on 28 February 2007; the FOV is $56'' \times 56''$.
}
\label{mfgs-High}
\end{figure}

The images with the highest value in each data set are shown in Figure~\ref{mfgs-High} as labeled ``a'', ``b'', ``c'', and ``d'', respectively. One can see clearly that the change in their image quality agrees with their MFGS value well. The image taken with {\it Hinode}/SOT is also shown in Figure~\ref{mfgs-High}e. Its image quality is much better than the others and its MFGS value (0.984) is much higher.

\begin{figure}    
\centerline{\includegraphics[width=0.8\textwidth,clip=]{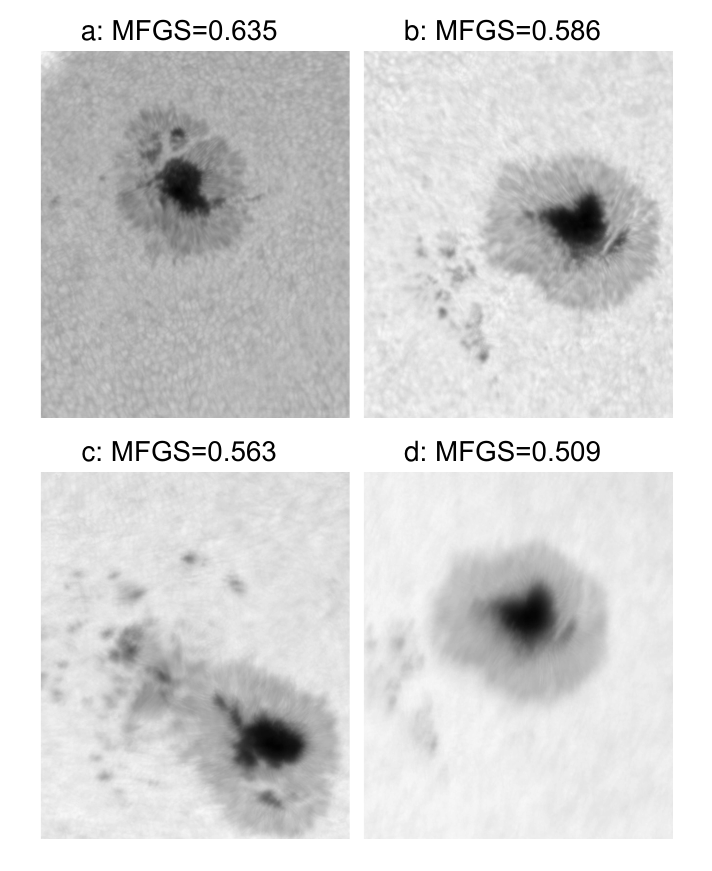}}
\caption{%
The images with the lowest MFGS value from data sets A, B, C, and D.
}
\label{mfgs-Low}
\end{figure}

Next, the images with the lowest value in each data set are shown in Figure~\ref{mfgs-Low} as labeled ``a'', ``b'', ``c'', and ``d'', respectively. It is obvious that the change in their image quality also agrees with their MFGS value, and the overall quality of these images is much worse than the images in Figure \ref{mfgs-High}.

\section{Comparison between RMS-Contrast and MFGS}
\label{S-Comparison}
In this section, MFGS and RMS-contrast are compared, particularly by paying attention to their dependence on the image properties. For this purpose, each image in the data sets given in Table~\ref{T-dataset} was divided into two parts, {\it i.e.}, the active region area and quiet-Sun region area; the latter included no sunspots nor pores. Then MFGS and RMS-contrast were applied separately to these two areas. The RMS-contrast value is defined as
\begin{equation}
\textrm{rms-contrast}=\sqrt{\frac{1}{N}\sum_{i=1}^{N} \left(I-\bar{I}\right)^{2}}/\bar{I},
\end{equation}
where $\bar{I}$ is the mean intensity and $N$ is the number of pixels~\citep{Roudier1986}. The higher the quality of the image is, the bigger the value of the contrast would be.

\begin{figure}    
\centerline{\includegraphics[width=1\textwidth,clip=]{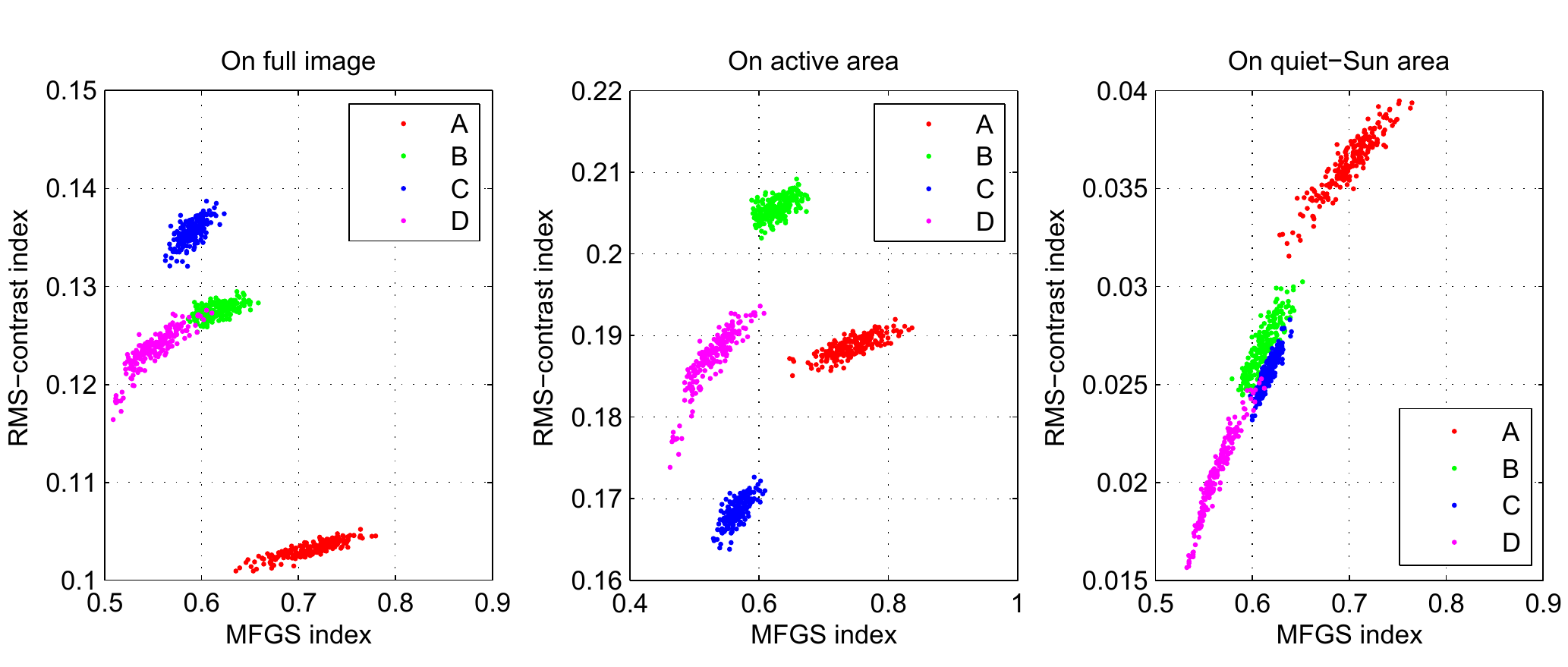}}
\caption{%
Scatter plots between the  RMS-contrast and MFGS metrics. The left panel shows the results from full images; the middle and the right panels show the results from active region areas and quiet-Sun areas, respectively. Red, green, blue, and magenta indicate data sets A, B, C, and D.
}
\label{MFGSvsRMS4}
\end{figure}

\subsection{Correlation between MFGS and RMS-Contrast}
\label{s-Correlation analysis}
Figure~\ref{MFGSvsRMS4} shows the correlation between MFGS and RMS-contrast metrics derived from the four image sets. The left panel shows that the images with good quality in data set A are erroneously determined as bad ones by the RMS-contrast metric. The middle panel shows that the images in data set B are erroneously determined as good ones by the RMS-contrast metric. However, the right panel displays a good linear correlation between MFGS and RMS-contrast metrics, which means that RMS-contrast and MFGS are tightly linearly correlated when performed on quiet-Sun structures (granules) with uniform and isotropic characteristics. Moreover, the variation in the RMS-contrast or MFGS values in the quiet-Sun areas agrees with the $r_{0}$ values better than those in the full images or active region areas.

Accordingly, we calculated the correlation coefficients between the MFGS and RMS-contrast metrics (Table~\ref{T-Correlation}). The correlation coefficients for the quiet-Sun area ({\it i.e.}, 0.932, 0.889, 0.902, 0.968) are higher than those for the full images or for active region areas. This also confirms the result shown in Figure~\ref{MFGSvsRMS4}, {\it i.e.}, a linear correlation between MFGS and RMS-contrast exists when performed on the quiet-Sun areas.

\begin{table}
\caption{%
Correlation coefficients between MFGS and RMS-contrast metrics.
}
\label{T-Correlation}
\begin{tabular}{ccccc}
\hline
                &A       &B       &C       &D       \\
\hline
Full image      &0.862  &0.571  &0.692  &0.867   \\
Active region area   &0.769  &0.571  &0.739  &0.878   \\
Quiet-Sun area &0.932  &0.889  &0.902  &0.968   \\
\hline
\end{tabular}
\end{table}

\subsection{Dependence on Image Contents}
\label{s-Content dependence}
A good image quality metric should perform well, independent of image contents. In our case, the performance of a metric on the active region areas and the quiet-Sun areas can give some hint on this point. Therefore, we compared the metric values derived from the two areas. A weak correlation between the two values implies that the corresponding metric heavily depends on the image contents, and vice versa. Table~\ref{T-dependence} shows the results for the MFGS and RMS-contrast metrics. For the RMS-contrast, the correlation coefficients are below 0.6 for data sets B and C, which indicates that the performance of the RMS-contrast metric depends significantly on the image contents. A photospheric image contains a variety of structures including granules, intergranular dark lanes, bright points, pores, and sunspots; rich in image contents. For the active region areas that contain pores and sunspots, the RMS-contrast value is dominated by the intensity differences among umbra, penumbra, and granules. With dark features (sunspots) moving in and out of the FOV, the RMS-contrast value will change significantly. For MFGS, the correlation coefficients are between 0.978 (maximum) and 0.700 (minimum), which indicates that the MFGS metric depends less on the image contents compared with the RMS-contrast metric. From this point of view, MFGS is better than RMS-contrast.

\begin{table}
\caption{%
Correlation coefficients between the metric values obtained from active region areas and quiet-Sun areas by MFGS and RMS-contrast.
}
\label{T-dependence}
\begin{tabular}{ccccc}
\hline
                &A       &B       &C       &D       \\
\hline
MFGS           &0.823 &0.798 &0.700  &0.978     \\
RMS-contrast   &0.758 &0.567 &0.547  &0.896      \\
\hline
\end{tabular}
\end{table}

\section{Discussion and Conclusions} 
\label{S-ConclusionDiscussions}
We propose a new NR/blind objective image quality metric, {\it i.e.}, the MFGS metric. The performance tests on short-exposure photospheric images acquired by NVST and the images acquired by {\it Hinode}/SOT show that the MFGS values change monotonically from 1 to 0 with the degradation of image quality. There exists a linear correlation between the derived values of MFGS and granular RMS-contrast. Moreover, MFGS is less affected by the image contents than RMS-contrast and it is a good alternative for the quality assessment of photospheric images.

There are issues that are worthy of discussion. First, all of the short-exposure sample images used in this study were obtained with NVST without its AO system. Although speckle image techniques have been successful in obtaining diffraction limited images, the most powerful technique for large astronomical telescopes is the one using AO technique. Therefore, an IQM is required to quantify the level of seeing compensation by the AO system. For MFGS is a good measure of photospheric image quality, we believe that it may be used in AO systems, but further investigation would be needed to validate it. Second, considering the fundamental principle of MFGS, we believe that it could also be used to assess images with rich structural information and sharp edges, like high resolution chromospheric images, but the expected performance should be checked through further studies.

\section*{Acknowledgements}
This work is supported by the National Natural Science Foundation of China (11163004, U1231205, 11263004, 11303011, 11103005, 11463003 and 11203077) and Natural Science Foundation of Yunnan Province (2013FA013, 2013FA032, 2013FZ018 and 2013FZ018). Wenda Cao acknowledges the support of the US National Science Foundation (AGS-0847126, AGS-1146896). {\it Hinode} is a Japanese mission developed and launched by ISAS/JAXA, with NAOJ as domestic partner and NASA and STFC (UK) as international partners. It is operated by these agencies in co-operation with ESA and NSC (Norway). We thank the referee for providing valuable suggestions which substantially helped to improve the quality of the paper.

\bibliographystyle{spr-mp-sola}

\tracingmacros=2
\bibliography{ImageAssess_ansi}

\IfFileExists{\jobname.bbl}{} {\typeout{}
\typeout{****************************************************}
\typeout{****************************************************}
\typeout{** Please run "bibtex \jobname" to obtain} \typeout{**
the bibliography and then re-run LaTeX} \typeout{** twice to fix
the references !}
\typeout{****************************************************}
\typeout{****************************************************}
\typeout{}}
\end{article}

\end{document}